\documentclass[12pt,reqno]{article}
\PassOptionsToPackage{linktocpage}{hyperref}
\usepackage[numbers,sort&compress]{natbib}
\usepackage{setspace}
\usepackage{xfrac}
\usepackage{fullpage}
\usepackage{physics}
\usepackage{amsmath}
\usepackage{tikz}
\usepackage{mathdots}
\usepackage{yhmath}
\usepackage{cancel}
\usepackage{color}
\usepackage{siunitx}
\usepackage{array}
\usepackage{multirow}
\usepackage{caption}
\usepackage{amssymb}
\usepackage{array}
\usepackage{textcomp,gensymb}
\usepackage{makecell}
\usepackage{makecell}
\usepackage{float}
\usepackage{cancel}
\usepackage{longtable}
\usepackage{tabularx}
\usepackage{booktabs}
\usetikzlibrary{fadings}
\usepackage{graphicx}
\usepackage{float}
\usepackage[hidelinks]{hyperref}
\usepackage{comment}
\usepackage{physics}
\usepackage{amsmath}
\usepackage{tikz}
\usepackage{mathdots}
\usepackage{yhmath}
\usepackage{cancel}
\usepackage{color}
\usepackage{siunitx}
\usepackage{array}
\usepackage{epsfig}
\usepackage{multirow}
\usepackage{amssymb}
\usepackage{gensymb}
\usepackage{booktabs}
\usetikzlibrary{fadings}
\usepackage{tensor}
\usepackage{ltablex}
\usetikzlibrary{patterns}
\usetikzlibrary{shadows.blur}
\usetikzlibrary{shapes}
\usepackage{appendix}
\usepackage{mdframed}
\newcommand{\be}{\begin{equation}}
\newcommand{\ee}{\end{equation}}
\newcommand{\nn}{\nonumber}
\newcommand{\ba}{\begin{eqnarray}}
\newcommand{\ea}{\end{eqnarray}}

\begin{document}
\begin{titlepage}
\vspace*{-10mm}
\hfill \vbox{
    \halign{#\hfil         \cr
           RCHEP/23-004\cr
           CERN-TH-2023-199\cr
           } 
      }  

\hfill \\
\vspace*{15mm}
\begin{center}
{\Large \bf Confinement from Distance in Metric Space and its Relation to Cosmological Constant} 

\vspace*{15mm}

{\large $\text{Amineh Mohseni}^{a,b,}\footnote{\href{mailto:amineh.mohseni@cern.ch}{amineh.mohseni@cern.ch}}~\text{, Mahdi Torabian}^{a,c,}\footnote{\href{mahdi.torabian@sharif.ir}{mahdi.torabian@sharif.ir}} $}
\vspace*{6mm}

\centerline{\it $^a$Center for High Energy Physics, Department of Physics, Sharif University of Technology, Tehran, Iran} 
\vspace{0.1cm}\centerline{\it $^b$CERN, Theoretical Physics Department, 1211 Meyrin, Switzerland}
\vspace{0.1cm}\centerline{\it $^c$Perimeter Institute for Theoretical Physics, Waterloo, ON, N2L 2Y5, Canada}

\vspace*{0.7cm}


\end{center}

\begin{abstract}
	We argue that, in a theory of quantum gravity, the gauge coupling and the  confinement scale of a gauge theory are related to distance in the space of metric configurations, and in turn to the cosmological constant. To support the argument, we compute the gauge kinetic functions in variuos supersymmetric Heterotic and type II string compactifications and show that they depend on distance. According to the swampland program, the distance between two (anti) de Sitter vacua in the space of metric configurations is proportional to the logarithm of the ratio of cosmological constants and thus the confinement scale depends on the value of the cosmological constant. 
    In this framework, for de Sitter space, we revisit the swampland Festina Lente  bound and gauge theories in the dark dimension scenario. 
    We show that if the Festina Lente bound is realized in a de Sitter vacuum and dependence on distance is strong enough, it will be realized in vacua with higher cosmological constants. In dark dimension scenario, as the value of cosmological constant is related to the decompactifying dimension, we find that the confinement scale is indeed related to radius of dark dimension. We show that in this scenario the Festina Lente bound holds for the standard model QCD, as well as all confining gauge groups with $N_c\lesssim 10^3$.
\end{abstract}

\end{titlepage}


\section{Introduction}\label{sec1}

The swampland program puts forward the idea that not all self-consistent low energy effective theories (EFTs) can be UV completed in a theory of quantum gravity. Its goal is to determine the set of conditions an EFT consistent with quantum gravity must satisfy \cite{Vafa:2005ui,Ooguri:2006in}. These criteria are formulated as conjectures with different levels of rigorousness (see \cite{Brennan:2017rbf,Palti:2019pca,vanBeest:2021lhn,Agmon:2022thq} for review).  The swampland program provides guidelines for low energy physics, leads to explaining the properties of the observed universe and genuine predictions. 
It explains properties of low energy physics and in particular the value of low energy parameters which may otherwise seem unnatural ({\it i.e.} the Higgs mass parameter and the cosmological constant) \cite{Gonzalo:2021zsp,Ibanez:2017kvh,Ibanez:2017oqr,montero2022dark}.

Among well-established conjectures are no global symmetries \cite{banks2011symmetries}, the swampland distance conjecture (SDC) \cite{ooguri2007geometry} and weak gravity conjecture (WGC) \cite{arkani2007string} which constitute core of the swampland program. These conjectures are not very predictive for low energy physics though. However, recently proposed conjectures including the swampland de Sitter conjecture  \cite{obied2018sitter,ooguri2019distance}, the trans-Planckian censorship conjecture (TCC) \cite{bedroya2020trans} and the Festina Lente (FL) bound \cite{Montero:2019ekk,Montero:2021otb} have direct implications for the low energy EFT. An interesting observation is that the swampland conjectures are related to each other.
It suggests that they might be different aspects of quantum gravity principles and thus 
it is important, in the swampland program, to find out how different conjectures are related. Specially, connections between the predictive conjectures and the well-established ones would be interesting.

Yang-Mills gauge theories, characterized by a gauge group and a coupling constant, are indispensable parts of our description of low energy physics. 
In the framework of quantum field theory, the dimensionless gauge coupling in four dimensions is replaced by a confinement scale via dimensional transmutation. 
In the light of swampland program, it is interesting to study the confinement dynamics in a theory consistent with quantum gravity. 
The FL bound, motivated by studying black holes in dS space, puts a lower bound on mass of charged particles in dS space. In particular, it implies that a non-Abelian gauge theory in dS space must be either confined (at a scale higher or equal to the Hubble scale) or Higgsed. This condition is fulfilled in the Standard Model sector in the present dS phase of the universe; the electroweak symmetry is spontaneously broken and QCD is confined. However, in the absence of other dynamics, the FL bound becomes a non-trivial constraint in a dS phase with a higher cosmological constant {\it e.g.} during primordial inflation. 
In fact, the FL bound in different dS vacua is respected if the confinement scale increases as the cosmological constant increases. 

In this paper, to examine the above possibility, we study the connection between the confinement scale of a gauge theory and the SDC. According to the generalized distance conjecture, the distance between vacua whose metrics are related by a Weyl transformation is the Weyl factor up to an order one constant \cite{lust2019ads}. We refer to it as \textit{distance in the space of metric configurations} in the rest of the paper.  We study the dependence of the gauge coupling on distance in the space of metric configurations and its implications for the confinement scale. To be specific, we study semi-realistic compactifications of heterotic, type IIA, and type IIB string theories in which we can compute the gauge kinetic functions in four dimensions. We find that the gauge coupling actually depends on distance in the space of metric configurations which consequently implies a distance dependent confinement scale.  We expand the gauge kinetic function for small changes of distance so that we can draw general conclusions  regardless of the exact form of the distance dependence.  For (Anti) dS space, distance in the space of metric configurations is proportional to the logarithm of the cosmological constant. Therefore, our result implies that the confinement scale of a Yang-Mills theory depends on the cosmological constant. We find the condition that the FL bound is respected in all dS space with different vacuum energy. Furthermore, the arguments establish a relation between the SDC and the FL bound. Interestingly, our result is generic and is independent of 
the exact form of the scalar potential in the EFT or the details of moduli stabilization. Although the computable examples of compactifications we used are supersymmetric, we make general predictions which is believed to be held in non-supersymmetric constructions. The bottomline is that stringy effects are stronger that running due to loop effects.

Moreover, we study the distance dependent confinement scale and the FL bound in the dark dimension scenario which ties the dS space with small cosmological constant to decompactification of one extra dimension \cite{montero2022dark}. It is motivated by the observation of a tiny cosmological constant which indicates our universe sits in asymptotic limit of moduli space. In this scenario, the vacuum energy is sourced by the zero-point energy of lightest Kaluza-Klein modes and the Standard Model fields are localised on a brane perpendicular to the dark dimension. We study F-theory realization of this scenario with generic $CY_4$ compactification and GUT $D_7$ brane.
In this setup we find that the gauge decreases (equivalently the confinement scale increases) as the radius of dark dimension decreases ({\it i.e.} cosmological constant increases). It is in agreement with the FL bound for confining gauge theories. 

The structure of the paper is as follows; in section 2 we generally study a gauge theory with distance-dependent gauge kinetic function and derive the confinement scale as a function of distance in the metric space. We present examples of string theory compactifications to four dimensions. In section 3 we look into the implications of distance-dependent gauge coupling and the confinement scale in the (A)dS space and we find the condition that the FL bound is realized in all dS vacua. We also look at the dark dimension scenario for dS space with positive vacuum energy show that how the FL bound is respected in this scenario. Finally, we conclude in section 4.

\section{Distance-Dependent Confinement Scale}
According to generalised distance conjecture a distance can be associated to any dynamical field $\mathcal{O}_{I_1,\dots I_n}$ as
\be \Delta_{\mathcal{O}}=\int \left(K^{I_1,\dots I_n,J_1,\dots J_n} \,\frac{\partial\mathcal{O}_{I_1,\dots I_n}}{\partial\tau}\frac{\partial\mathcal{O}_{J_1,\dots J_n}}{\partial\tau}\right)^{\sfrac{1}{2}}d\tau,\ee
where $K$ is the metric in field space of $\mathcal{O}$, and $\tau$ parameterizes the geodesic path in field space.
In particular, it is conjectured that the distance between vacua (or geometries) whose metrics are related via the following \textit{external} Weyl transformation
\be \tilde{g}_{\mu\nu}= e^{2 \Phi} g_{\mu \nu}, \ee
is given by field $\Phi$ (the Weyl factor) up to an order one constant, k, namely
\be \Delta = k \,\Phi,\label{delta} \ee
where it is proposed that, due to various contributions, the coefficient is such that the distance is  positive and well-defined \cite{lust2019ads}. If one only considers variation of the conformal factor, the distance turns out to be negative. This is the  well-known conformal problem. In a recent study \cite{li2023towards}, it has been shown that incorporating other contributions from string theory, namely variations of internal dimensions and fluxes, results in a positive distance. There have also been attempts to calculate the distance between flux vacua using open string moduli from D-branes interpolating between them \cite{shiu2023ads,shiu2023connecting,farakos2023off,tringas2023anisotropic,farakos2023scale}. In this work, we accept the (generalized) AdS distance conjecture as given.

In this paper we argue that due to quantum gravity effects the gauge kinetic function of a Yang-Mills theory typically depends on distance in the space of  metric configurations, $\Phi$. To support our argument, in the next section, we  present examples from string theory. In this section, we study the effect of this dependence on the confinement scale of a Yang-Mills theory given the action
\ba  \mathcal{L}_{\text{YM}}&=& - \frac{1}{4} f(\Phi)\, {\rm tr}\, (F_{\mu \nu}F^{\mu\nu}), \ea
 where $\Phi$ measures distance between vacua in the space of metric configurations. If we move in the space of metric configurations from a point where the UV (string scale) gauge coupling is $g_{0,\rm UV}$ (call it $\Phi=0$), the gauge coupling at a point at distance $\Phi$ from the initial point would be
\be g_{\rm UV}^{-2}=g_{0,\rm UV}^{-2}+[ f(\Phi)-f(0) ].\ee
 Furthermore, the gauge coupling runs with energy due to radiative corrections. Therefore, the gauge couplings in two vacua in the IR are related as
\be \label{gauge coupling2} \alpha_{\rm IR}(\Phi,\mu)^{-1}=\alpha_{\rm IR}(0,\mu_0)^{-1}+\frac{\beta_0}{4\pi} \ln \frac{\mu^2}{\mu_0^2} +4 \pi [f(\Phi)-f(0)],\ee
where $\alpha_{\rm IR}\equiv\frac{g_{\rm IR}^2}{4\pi}$ and $\beta_0$ is one-loop beta coefficient (for tha sake of simplicity, we assume that two theories in each vacuum we are comparing have similar field contents). 
If the gauge theory confines at $\mu_{\text{conf}}(0)$ when $\Phi=0$, then, at distance $\Phi$ away from that point the confinement scale would be 
\be \label{fielddep confinement scale2}   \mu_{\text{conf}}(\Phi)=\mu_{\text{conf}}(0) \,e^{-\frac{8 \pi^2}{\beta_0}[f(\Phi)-f(0)]},\ee
thus the confinement scale has a distance dependence induced by the gauge kinetic function.
 
Our goal is to make a general arguments from this observation regardless of the exact form of the gauge kinetic function or underlying dynamics of moduli stabilization. This can be done if we consider small changes in distance and consider the leading order expansion of the gauge kinetic function given $|\Phi|<1$ as \footnote{Note that working with $|\Delta\Phi|\equiv|\Phi|<1$ is not in contradiction with working in the asymptotic limit, which is when $|\Phi_0|>1$. } 
 \ba \label{gauge coupling} f \simeq f|_{\Phi=0}+\left(\partial_{\Phi}f\right)|_{\Phi=0} \Phi \equiv  g_{0,\rm UV}^{-2}(1+r\Phi),\ea
where $f|_{\Phi=0} = g_{0,\rm UV}^{-2}$ is the UV (string scale) gauge coupling at $\Phi=0$ and $r$ is defined as
\be  \label{rcalc}r=\frac{\left(\partial_{\Phi}f\right)|_{\Phi=0}}{f|_{\Phi=0}}, \ee
we drop the UV index in the rest of the paper. Then, including the loop effects, confinement scale as a function of distance is
 \be \label{confinementLambda} \mu_{\text{conf}}(\Phi)=\mu_{\text{conf}}(0) \, e^{-\frac{8 \pi^2}{g_0^2\beta_0}r\Phi}.\ee
 For non-zero positive or negative $r$, the gauge coupling respectively decreases or increases as distance in metric space increases. 

\subsection{Evidence from String Theory}
In this section, to support our argument, we present examples of four dimensional compactifications of string theory where  gauge couplings depend on distance in the space of metric configurations. In particular, we study  compactifications of the heterotic, type IIA, type IIB with $D7$/$D3$ branes and type IIB with $D9/D5$ branes (type I) to four dimensions. 

The (relevant part of) string frame action after compactification to four dimensions is
\be\label{stringframe4d} S\supset \frac{1}{2 \alpha^{\prime}} \int d^{4} x \sqrt{-G^s} \,e^{-2 \phi_4} \left(R^s+\Lambda^s\right),\ee
where $\phi_4$ is the four dimensional dilaton, which is a function of the ten-dimensional dilaton and the compactification volume ${\cal V}$ 
\be \phi_4=\phi-\frac{1}{2}\ln\left({\cal V}/\alpha^{\prime\,3}\right) .\ee
We consider two vacua in the space of metric configurations which are distinguished with their string frame metrics $G^{s,(1)}_{\mu\nu}$ and $G^{s,(0)}_{\mu\nu}$, and their four dimensional dilaton $\phi_4^{(1)}$ and $\phi_4^{(0)}$. As we eventually apply our result to (A)dS background, we consider metrics that differ by a Weyl transformation as
\be G_{\mu\nu}^{s,(1)}=e^{-2\omega} G_{\mu\nu}^{s,(0)}, \ee
where $\omega$ is in general, a function of ten dimensional dilaton and volume moduli. 

The action can be written in the Einstein frame through a Weyl transformation of the metric 
\be g^{E,(i)}_{\mu\nu}=e^{-2 (\phi^{(i)}_4-c^{(i)})} G^{s,(i)}_{\mu\nu},\ee
where $i=0,1$ and $c^i$ is a constant. We fix the Einstein frame by choosing $c^{(0)}=c^{(1)}=-\sfrac{1}{2} \ln(2 \alpha^{\prime})$ which sets the four dimensional Planck mass to one ($m_{{\rm Pl},4}=1$)
at both points in metric space. It guarantees that we have removed Weyl redefinition without physical meaning given by $c^{(0)}\neq c^{(1)}$. Then, the \textit{External} metrics of the two vacua we compare  are related through the following Weyl transformation in Einstein frame
\be g^{E,(1)}_{\mu \nu}=e^{-2(\phi_4^{(1)}-\phi_4^{(0)}+\omega)} g^{E,(0)}_{\mu \nu} ,\ee
Moreover, the cosmological constants are related as
\be \Lambda^{(1)}=e^{2(\phi_4^{(1)}-\phi_4^{(0)}+\omega)} \Lambda^{(0)}. \ee
Finally, according to generalized distance conjecture, the distance between these vacua is given by
\be \label{distance} \Phi=-(\phi_4^{(1)}-\phi_4^{(0)}+\omega), \ee
up to an order one constant. 

The gauge kinetic function of a Yang-Mills theory obtained from string compactifications to four dimensions typically depends on field $\Phi$ through its dependence on volume and ten-dimensional dilaton. In fact, we show that $\partial_\Phi f\neq0$ by computing
 \ba \label{partialder}\partial_{\Phi}f=\sum_{i}\frac{\partial f}{\partial \rho_i} \frac{\partial \rho_i}{\partial \Phi}+\frac{\partial f}{\partial \phi} \frac{\partial \phi}{\partial \Phi}, \ea
 where $\rho_i$ are the geometric moduli of the compact manifold and $\phi$ is the dilaton. The partial derivatives $\sfrac{\partial \rho_i}{\partial \Phi}$ and $\sfrac{\partial \phi}{\partial \Phi}$ tell us how moduli $\rho_i, \phi$ change (through tunneling or rolling). These derivatives are determined by the structure of vacua or dynamics of the underlying quantum gravity theory, and account for the change of $\Phi$. 

For the sake of concreteness, we study compactification on factorizable $T^6$ which allows explicit calculations.  Although toroidal compactification is supersymmetric, we expect that our main point which is dependence of the gauge kinetic function on volume moduli and ten dimensional dilaton and therefore distance between vacua, holds also for compactification on more general manifolds. For factorizable $T^6$, equation \eqref{partialder} is computed as
\ba \label{chain-explicit} \partial_{\Phi}f&=&\sum_{i=1}^{3}\frac{\partial f}{\partial R_x^i}\frac{\partial R_x^i}{\partial  \Phi} + \sum_{i=1}^{3}\frac{\partial f}{\partial R_y^i}\frac{\partial R_y^i}{\partial \Phi} + \frac{\partial f}{\partial \phi}\frac{\partial \phi}{\partial \Phi}, \ea
where $R_x^i$ and $R_y^i$ are dimensionless radii in units of $\alpha^{\prime}$.

In order to compute the above constrained partial derivative, we parameterize change of the radii and the ten dimensional dilaton in the following way; such that $g_i$ contain information on how the geometric moduli change due to dynamics (tunneling or rolling) of the underlying theory 
\ba &&{\rm d} R_x^i\equiv g^i_x {\rm d}\Phi,\\
&&{\rm d} R_y^i\equiv g^i_y {\rm d}\Phi,\\
&&{\rm d} \phi\equiv g_{\phi} {\rm d}\Phi,\ea
change of different moduli accounts for the change of distance, that is the dynamics is subject to the constraint
\be \label{const1} \Phi=-(\phi_4-\phi_4^0+\omega),  \ee 
where $\omega$ is a function of the ten dimensional dilaton and the geometric moduli, $\omega(R_x^i,R_y^i,\phi)$, and the four dimensional dilaton in terms of geometric data is 
\be \phi_4=\phi-\frac{1}{2} \ln\big(\prod_{i=1}^{3}R_x^i R_y^i \big).  \ee 
Upon differentiating, the above constraint implies 
\ba g_{\phi}=\frac{1}{2}\frac{1}{1+\frac{\partial \omega}{\partial \phi}} \Big[\sum_{i=1}^{3}\frac{g^i_x}{R_x^i}\Big(1- 2\frac{\partial \omega}{\partial R_x^i} R_x^i\Big)
+\sum_{i=1}^{3}\frac{g^i_y}{R_y^i}\Big(1-2\frac{\partial \omega}{\partial R_y^i} R_y^i\Big)-2\Big].\quad \ea
The derivatives $\sfrac{\partial \omega}{\partial R_x^i},\sfrac{\partial \omega}{\partial R_y^i},\sfrac{\partial \omega}{\partial \phi}$ contain information on details of the compactification, which we leave as general functions. Substituting the above relation for $g_{\phi}$ in \eqref{chain-explicit}, we get
\ba \partial_{\Phi}f&=&\sum_{i=1}^{3}\left( \frac{\partial f}{\partial R_x^i}+\frac{1}{2R_x^i}\frac{1- 2\frac{\partial \omega}{\partial R_x^i} R_x^i}{1+\frac{\partial \omega}{\partial \phi}}\frac{\partial f}{\partial \phi}\right) g^i_x\cr
&+&\sum_{i=1}^{3}\left( \frac{\partial f}{\partial R_y^i}+\frac{1}{2R_y^i}\frac{1- 2\frac{\partial \omega}{\partial R_y^i} R_y^i}{1+\frac{\partial \omega}{\partial \phi}}\frac{\partial f}{\partial \phi}\right) g^i_y\cr
&-&\frac{1}{1+\frac{\partial \omega}{\partial \phi}}\frac{\partial f}{\partial \phi}, \ea
the ratio of expansion coefficients is  then straightforwardly calculated according to \eqref{rcalc}. 
The properties of the structure of vacua or dynamics of the underlying theory is captured by $r$
\subsubsection{Heterotic String Compactifications}
Consider the low energy effective theory for compactifications of  heterotic string. Symmetry arguments,  \cite{polchinski2005string}, imply that gauge kinetic function of gauge fields in the effective four dimensional theory is as follows
\ba &&f_{ab}=S \delta_{ab},\\
 && S=e^{- 2 \phi_4}+i a,\ea
ratio of the expansion coefficients, which is a measure of distance dependence is
\ba r_{\text{het}}= \frac{1}{1+\frac{\partial \omega}{\partial \phi}}\Big[2+\sum_{i=1}^{3}\frac{g_x^i}{R_{x}^i}\Big(\frac{\partial \omega}{\partial \phi}+2R_{x}^i \frac{\partial \omega}{\partial R_x^i}\Big)+\sum_{i=1}^{3}\frac{g_y^i}{R_{y}^i}\Big(\frac{\partial \omega}{\partial \phi}+2 R_{y}^i\frac{\partial \omega}{\partial R_y^i}\Big)\Big],\ea
where the radii and derivatives are evaluated at the expansion point. We note that $r_{\text{het}}$ captures the parameters $g_x^i, g_y^i$ which contain information on how the geometric moduli change, also derivatives of the function $\omega$ that contain information about the details of the compactification, and the vacuum. This is also the case for type II examples that we investigate bellow.

\subsubsection{Type II Compactifications}
The gauge coupling of the gauge symmetry resulting from a stack of D-branes in type II compactifications follows from expanding the DBI action to quadratic order in gauge field strength. For $D_p$ brane the gauge coupling is as follows
 \be \label{type2coupling} f_p=e^{-\phi} \frac{\alpha^{\prime\,\sfrac{(3-p)}{2}}}{(2 \pi)^{p-2}}  \text{Vol}(\Pi_{p-3}),\ee
where $f_p\equiv\sfrac{1}{g_p^2}$, $\text{Vol}(\Pi_{p-3})$ is volume of the cycle wrapped by the brane, and $\phi$ is the ten dimensional dilaton. The gauge coupling \eqref{type2coupling} has been calculated in terms of geometric data for compactification on a factorizable $T^6$ in \cite{ibanez2012string}.

Firstly, we consider type IIA with D6 branes.
The gauge kinetic function is
\be f_6=\frac{e^{-\phi}}{2 \pi} \prod_{i=1}^{3}\left(n_i^2 R_x^{i\,2}+m_i^2 R_y^{i\,2} \right) ^{\sfrac{1}{2}},\ee
where $(m_i, n_i)$ are the wrapping numbers.  The ratio of expansion coefficients is
\ba r_6=\frac{1}{1 + \frac{\partial \omega}{\partial \phi}}\Big[1\!\!\!\!&+&\!\!\! \frac{\partial \omega}{\partial \phi}\sum_{i=1}^{3}\frac{n_i^2 g_x^i R_{x}^i+m_i^2 g_y^i R_{y}^i}{n_i^2 (R_{x}^i)^2+m_i^2 (R_{y}^i)^2}\cr
&+&\!\!\! \frac{1}{2}\sum_{i=1}^{3} \frac{g_x^i}{R_{x}^i}\frac{n_i^2 (R_{x}^i)^2(1+ 2R_{x}^i \frac{\partial \omega}{\partial R_x^i})-m_i^2 (R_{y}^i)^2(1- 2R_{x}^i \frac{\partial \omega}{\partial R_x^i})}{n_i^2 (R_{x}^i)^2+m_i^2 (R_{y}^i)^2} \cr
&+&\!\!\! \frac{1}{2}\sum_{i=1}^{3} \frac{g_y^i}{R_{y}^i}\frac{m_i^2 (R_{y}^i)^2(1+ 2 R_{y}^i\frac{\partial \omega}{\partial R_y^i})-n_i^2 (R_{x}^i)^2(1- 2 R_{y}^i\frac{\partial \omega}{\partial R_y^i})}{n_i^2 (R_{x}^i)^2+m_i^2 (R_{y}^i)^2} \Big].\quad\nn\\\ea

Secondly, we consider type IIB with D7/D3 branes.  The gauge kinetic functions are
\ba&& f_7^i=\frac{e^{-\phi}}{2 \pi} \prod_{j\neq i}\left(n_j^2 +m_j^2 R_x^{j\,2}R_y^{j\,2} \right) ^{\sfrac{1}{2}},\\
 &&f_3=\frac{e^{-\phi}}{2 \pi} ,\ea
where $m_i $, and $n_i$ are the wrapping numbers, and magnetic fluxes respectively. The ratio of the expansion coefficients for D7 brane is
\ba \label{D7gen}  r^i_7=\frac{1}{1+\frac{\partial \omega}{\partial \phi}}\Big[1\!\!\!&-&\!\!\!\frac{1}{2}\frac{g_x^i}{R_{x}^i}\Big(1-2R_{x}^i\frac{\partial \omega}{\partial R_x^i}\Big)-\frac{1}{2}\frac{g_y^i}{R_{y}^i}\Big(1-2R_{y}^i\frac{\partial \omega}{\partial R_y^i}\Big)\cr
&+&\!\!\! \frac{\partial \omega}{\partial \phi}\sum_{j\neq i}\frac{m_j^2 g_x^j R_{x}^j (R_{y}^j)^2}{n_j^2 +m_j^2 (R_{x}^j)^2(R_{y}^j)^2}+ \frac{\partial \omega}{\partial \phi}\sum_{j\neq i}\frac{m_j^2 g_y^j R_{y}^j (R_{x}^j)^2}{n_j^2 +m_j^2 (R_{x}^j)^2(R_{y}^j)^2}\cr
&+&\!\!\! \frac{1}{2}\sum_{j\neq i}\frac{g_x^j}{R_{x}^j}\frac{m_j^2 (R_{x}^j)^2(R_{y}^j)^2(1+ 2R_{x}^j\frac{\partial \omega}{\partial R_x^j})-n_j^2 (1- 2R_{x}^j\frac{\partial \omega}{\partial R_x^j})}{n_j^2 +m_j^2 (R_{x}^j)^2(R_{y}^j)^2}\cr
&+&\!\!\! \frac{1}{2}\sum_{j\neq i}\frac{g_y^j}{R_{y}^j}\frac{m_j^2 (R_{x}^j)^2(R_{y}^j)^2(1+ 2R_{y}^j \frac{\partial \omega}{\partial R_y^j})-n_j^2 (1- 2 R_{y}^j\frac{\partial \omega}{\partial R_y^j})}{n_j^2 +m_j^2 (R_{x}^j)^2(R_{y}^j)^2}\bigg].\ea
 The ratio takes a simpler form when there is no magnetization ($n_i=0$), in this case we have
\ba r^i_7=\frac{1}{1+\frac{\partial \omega}{\partial \phi}}\Big(1\!\!\!&-&\!\!\!\frac{1}{2}\frac{g_x^i}{R_{x}^i}
\Big(1-2R_{x}^i\frac{\partial \omega}{\partial R_x^i}\Big)-\frac{1}{2}\frac{g_y^i}{R_{y}^i}\Big(1-2R_{y}^i\frac{\partial \omega}{\partial R_y^i}\Big)\cr
&+&\!\!\! \frac{1}{2}\sum_{j\neq i}\frac{g_x^j}{R_{x}^j}(1+2\frac{\partial \omega}{\partial \phi}+2R_{x}^j\frac{\partial \omega}{\partial R_x^j})+\frac{1}{2}\sum_{j\neq i}\frac{g_y^j}{R_{y}^j}(1+2\frac{\partial \omega}{\partial \phi}+2R_{y}^j\frac{\partial \omega}{\partial R_y^j})\Big).\ea
For $D3$ brane, calculation of ratio of the expansion coefficients leads to the following result
\ba r_3=\frac{1}{1+\frac{\partial \omega}{\partial \phi}}\Big[1-\frac{1}{2}\sum_{i=1 }^{3}\frac{g_x^i}{R_{x}^i}\Big(1-2R_{x}^i\frac{\partial \omega}{\partial R_x^i}\Big) - \frac{1}{2} \sum_{i=1 }^{3}\frac{g_y^i}{R_{y}^i}\Big(1-2R_{y}^i\frac{\partial \omega}{\partial R_y^i}\Big)\Big].\ea

Finally, we consider type IIB with D9/D5 branes (type I). For simplicity we work with the case with no magnetization. The gauge kinetic functions are
\ba&& f_9=\frac{e^{-\phi}}{2 \pi} \prod_{i=1}^{3}R_x^{i}R_y^{i},\\
&&f_5^i=\frac{e^{-\phi}}{2 \pi} R_x^{i}R_y^{i}.\ea
The ratios of the expansion coefficients are 
\ba r_9=\frac{1}{1+\frac{\partial \omega}{\partial \phi}}\Big[1+\frac{1}{2}\sum_{i=1}^{3}\frac{g_x^j}{R_{x}^j}\Big(1+2\frac{\partial \omega}{\partial \phi}+2R_{x}^j\frac{\partial \omega}{\partial R_x^j}\Big)+\frac{1}{2}\sum_{i=1}^{3}\frac{g_y^j}{R_{y}^j}\Big(1+2\frac{\partial \omega}{\partial \phi}+2R_{y}^j\frac{\partial \omega}{\partial R_y^j}\Big)\Big],\  \ea
for $D_9$ brane, and
\ba r_5^i=\frac{1}{1+\frac{\partial \omega}{\partial \phi}}\Big[1\!\!\!&+&\!\!\!\frac{1}{2}\frac{g_x^i}{R_{x}^i}\Big(1+2\frac{\partial \omega}{\partial \phi}+2R_{x}^i\frac{\partial \omega}{\partial R_x^i}\Big)+\frac{1}{2} \frac{g_y^i}{R_{y}^i}\Big(1+2\frac{\partial \omega}{\partial \phi}+2R_{y}^i\frac{\partial \omega}{\partial R_y^i}\Big)\cr
&-&\!\!\!\frac{1}{2}\sum_{j\neq i}\frac{g_x^j}{R_{x}^j}\Big(1-2R_{x}^j\frac{\partial \omega}{\partial R_x^j}\Big) - 
\frac{1}{2}\sum_{j\neq i}\frac{g_y^j}{R_{y}^j}\Big(1-2R_{y}^j\frac{\partial \omega}{\partial R_y^j}\Big)\Big] ,\ea
for D5 brane.

We note that the order one parameter $r$, which indicates distance dependence, sums over changes of the geometric moduli 
\be r(g_x^i,g_y^i),\ee
 it can be positive or negative depending on the trajectory in the space of geometric moduli. We note that $r$ can also be zero in case the dynamics is tuned to move on special trajectories, which is improbable. For example in the case of type IIB with $D3$-branes, $r$ is zero if one moves on the trajectory 
 \be \sum_{i=1 }^{3}\frac{g_x^i}{R_{x}^i}\Big(1-2R_{x}^i\frac{\partial \omega}{\partial R_x^i}\Big)+\sum_{i=1 }^{3}\frac{g_y^i}{R_{y}^i}\Big(1-2R_{y}^i\frac{\partial \omega}{\partial R_y^i}\Big)=2.\ee
 This trajectory is a strong constraint requiring that the ten dimensional dilaton, and $\omega$ are the same fixed value during dynamics, and for all the vacua in case of tunneling. However, swampland program prefers exponentially decaying dynamics. We treat the order one parameter $r$ as a free parameter because we do not have information about dynamics or structure of vacua of the underlying theory. 
 
  As we mentioned before, although toroidal compactification is supersymmetric, the main point we hope to deliver is that the gauge kinetic function depends on volume moduli and ten dimensional dilaton which also enter vacuum energy and determine the distance. We expect that this feature also holds in case of compactification on more general manifolds, and after adding flux and stabilization.

\section{Application to (Anti) de Sitter Space}
The distance between two vacua whose metrics are related via a Weyl transformation is given by the Weyl factor. As argued in the previous sections, the gauge kinetic function and the confinement scale vary with distance in the space of metric configurations and their dependence is given by  \eqref{gauge coupling} and \eqref{confinementLambda} respectively. In this section, we apply our results to dS and AdS space. We derive dependence of the confinement scale on the cosmological constant only by referring to the metric of the (A)dS space and that  metrics of spaces with different cosmological constants are related to each other through a Weyl transformation. The arguments do not involve referring to the scalar potential or  stabilization problem, the validity of which is under debate for different constructions of dS (and AdS) space in string theory. We only refer to the metric of the space, and all the details about structure of vacua / dynamics of the underlying theory is captured by the parameter $r$.
 
In global coordinates four-dimensional dS and AdS metrics are
\ba  {\rm d}s_{\text{dS}}^2 &=& \frac{3}{|\Lambda|}\big(-{\rm d}t^2+\text{cosh}^2 t\, {\rm d}\Omega_3^2\big),\\
{\rm d}s_{\text{AdS}}^2 &=& \frac{3}{|\Lambda|}\big(-{\rm d}t^2+\text{sin}^2 t\left({\rm d} \psi^2+\text{sinh}\,\psi^2 {\rm d}\Omega_2^2\right)\big).\quad \ea
Apparently, a change of the cosmological constant is a Weyl transformation of the metric. Assume that along some path in the metric space the cosmological constant changes from $\Lambda_0$ to $\Lambda$. Then, according to generalized distance conjecture, the canonical scalar field measuring the distance between the (A)dS spaces is given by  field $\Phi$

 \be \label{distancelog}\Phi = k\, \ln \frac{\Lambda }{\Lambda_0} ,\ee
where $k$ is an order one constant,  taking all contributions to distance into account, this order one constant is such that the distance is positive \cite{lust2019ads}. As long as
 \be\label{limit} |\ln(\sfrac{\Lambda}{\Lambda_0})|\lesssim1,\ee 
 we have $|\Phi|<1$.

 As argued in the previous sections, to leading order the gauge kinetic function is
\ba \label{fexplicit} f(\Phi) \simeq \frac{1}{g_0^2} \Big(1+k r\ln \frac{\Lambda }{\Lambda_0}\Big),\ea
 one can see that the gauge coupling varies with the cosmological constant. Calculating the confinement scale from \eqref{confinementLambda}, we obtain
\be \label{conf in ds} \mu_{\text{conf}}(\Lambda)=\mu_{\text{conf}}(\Lambda_0) \left(\frac{\Lambda}{\Lambda_0}\right)^{\frac{-8 \pi^2 k r}{g_0^2\beta_0}}.\ee
 The above equation implies that the confinement has a power-law dependence on the cosmological constant of the (A)dS vacuum (in the regime of validity of our calculations). We find that for $\sfrac{k r}{\beta_0}>0$, $\sfrac{k r}{\beta_0}<0$ , or $\sfrac{k r}{\beta_0}=0$ the confinement scale decreases, increases, or remains constant respectively, as the cosmological constant increases. 

\subsection{Swampland Festina Lente Bound}
The Festina Lente (FL) bound is a lower bound on the mass of charged particles in dS space. Massless gluons charged under the Cartan subgroup of a non-abelian gauge group violate the Festina Lente bound. In other words, Nariai black holes can be embedded in the Cartan of a non-abelian gauge group. Requiring that these black holes decay without becoming super-extremal implies that non-abelian gauge fields in dS space must be confined at a scale above the Hubble scale if not spontaneously broken
\be\label{flconf2} \mu_{\text{conf}}\gtrsim H \sim \sqrt{\Lambda},\ee
where $H$ is the Hubble scale (see \cite{mishra2023confinement} for a related work). The FL bound also has implications for Higgs physics. The Higgs field must have a large vacuum expectation value during a primordial dS phase, assuming that it is the only source generating masses for particles (see \cite{Lee:2021cor,Mohseni:2022ftn}). Furthermore, the bound has implications for the hidden sector (see \cite{Ban:2022jgm,Guidetti:2022xct,Montero:2022jrc} for phenomenological studies.)

In the following we argue that {\it if} the bound \eqref{flconf2} is satisfied in a dS vacuum, it will also be satisfied in a vacuum with higher value of cosmological constant given that the stringy effects are large enough. This provides an insight on how the FL bound may be realised in string theory. It is especially important because even if  a Yang-Mills theory confines via radiative corrections in a certain vacuum such that the bound is satisfied, it is not guaranteed to be the case in a vacuum with higher cosmological constant. As an example consider QCD of strong interactions with confinement scale around 100 MeV. It satisfies the FL bound in the current vacuum by a huge margin or in a vacuum with positive cosmological constant up to $10^{-38}$ in Planck units. However, for higher cosmological constants (presumably during primordial inflation) the FL bound will not be satisfied if we only consider quantum effects. We will see  how the stringy effects studied in the previous section help to alleviate this problem via equation \eqref{conf in ds} which basically relates the FL bound to the generalised distance conjecture.

Assume that we move in the space of metric configurations from a point with positive cosmological constant $\Lambda_0$ to another point with a higher positive value of cosmological constant $\Lambda$. Then, \eqref{conf in ds} implies that the FL bound for confinement is
\be -16 \pi^2\frac{k r}{\beta_0 g_0^2} \ln \frac{\Lambda}{\Lambda_0}\gtrsim  \ln \frac{\Lambda}{\mu_{\text{conf}}(\Lambda_0)^2}. \ee
where we work in the regime given by \eqref{limit} for this result to make sense. From the above inequality follows that if the FL bound holds in the first vacuum, $\mu_{\rm conf}(\Lambda_0)\gtrsim\sqrt\Lambda_0$, then it will also be satisfied in the second vacuum, {\it i.e.}  
$\mu_{\text{conf}}(\Lambda)\gtrsim\sqrt\Lambda$, if 
\be \label{cnd2i}  -\frac{k r}{\beta_0 g_0^2}\gtrsim\frac{1}{16 \pi^2}. \ee
The above inequality states that if distance dependence of the gauge coupling, which is a stringy (quantum gravity) effect, is stronger than running due to loop effects the FL bound will hold in the second vacuum as well. We will look into this condition more precisely in case of the dark dimension scenario.

\begin{figure}[t]
	\centering
	\includegraphics[width=0.4\linewidth]{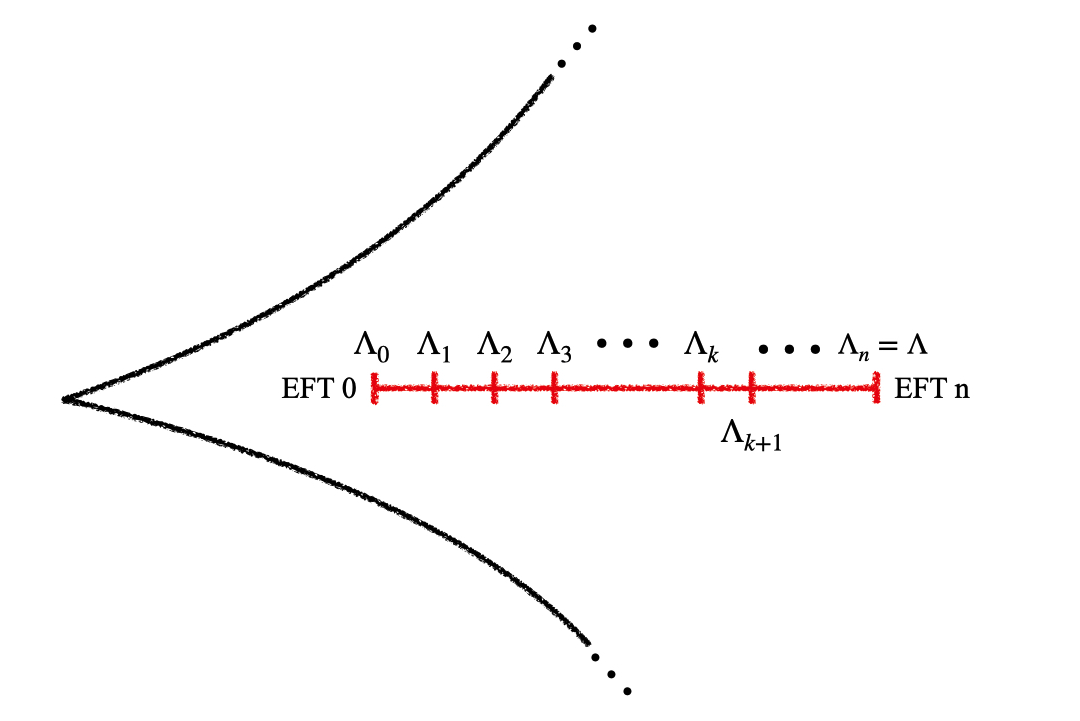}
	\caption{We can divide the path in the moduli space from $\Lambda_0$ to $\Lambda\gg\Lambda_0$ to many intervals each satisfy condition  \eqref{limit}}
	\label{fieldspace}
\end{figure}

The above argument can be extended to an arbitrary distance in the metric space; given that the FL bound is satisfied in a vacuum with cosmological constant $\Lambda_0$ we find the condition it is satisfied in any vacuum with an arbitrary higher cosmological constant $\Lambda_n = \Lambda$, if stringy effects are larger than loop effects. We consider a family of dS vacua parametrized by cosmological constant $\Lambda_i$, $i=0,1,\dots,n$ and assume $|\ln (\sfrac{\Lambda_{i+1}}{\Lambda_i})|\lesssim1$ for every $i$ (see figure \ref{fieldspace}) so that our expansion \eqref{gauge coupling} makes sense.
Given that $\mu_{\text{conf}}(\Lambda_i)\gtrsim \sqrt{\Lambda_i}$ holds, then for the $i$th interval
\be \mu_{\text{conf}}(\Lambda_{i+1})\gtrsim \sqrt{\Lambda_{i+1}},\ee
holds if 
\be \label{cnd2ii}-\frac{k_i r_{i}}{\beta_{0,i}g_{0,i}^2}\gtrsim\frac{1}{16 \pi^2}. \ee
Thus, if the FL bound is satisfied at point $\Lambda_0$ then through the generalized distance conjecture it is satisfied at every other point in the moduli space $\mu_{\text{conf}}(\Lambda_i)\gtrsim\sqrt{\Lambda_{i}}$, $i=1,\dots,n$.

\subsection{Dark Dimension Scenario}
Recently, it is argued that the present universe with $\Lambda\sim 10^{-122}$ is an asymptotic limit in the field space. According to the (A)dS distance conjecture a tower of light states with mass scale $ m\sim \Lambda^{\alpha}$ emerges in this limit where $\sfrac{1}{4}\leq\alpha\leq\sfrac{1}{2}$. The experimental and theoretical bounds imply that in the current universe
$\alpha=\frac{1}{4}$, and the tower is associated with decompactification of exactly one large extra dimension of radius 
\be \label{r} R\sim m^{-1} \sim \Lambda^{-\alpha},\ee
called dark dimension (see \cite{Gonzalo:2022jac,Anchordoqui:2023oqm,Anchordoqui:2022svl,Anchordoqui:2022tgp,Blumenhagen:2022zzw,Anchordoqui:2022txe,Anchordoqui:2022ejw} for related works and implications).

The canonical field that measures the distance between dS vacua with different radii of the dark dimension is given by \eqref{distancelog} so if the radius of the dark dimension changes from $R_0$ to $R$ then the distance is
\be \label{distancedarkdim} \Phi = -\frac{k}{\alpha} \ln \frac{R }{R_0} .\ee
We note that $\Phi\sim \ln{\cal V}\sim \ln R$ which is in agreement with the examples we considered from string theory.  As long as $ |\text{log}(\sfrac{R}{R_0})|\lesssim1$ we have $|\Phi|<1$ and the expansion \eqref{gauge coupling} can be trusted.
Then, to leading order the gauge kinetic function is
\ba f(\Phi) &\simeq& \frac{1}{g_0^2}\Big(1-\frac{k r}{\alpha}\ln\frac{R}{R_0}\Big). \ea
Apparently,  the gauge coupling changes with the radius of dark dimension and thus confinement scale is read as
\ba \mu_{\text{conf}}(R)&=&\mu_{\text{conf}}(R_0)\left(\frac{R}{R_0}\right)^{\frac{8\pi^2 k r}{\alpha g_0^2\beta_0}}.\ea
The confinement scale has power law dependence on the radius of dark dimension.

To be more specific about behaviour of the confinement scale as radius of the dark dimension changes, we must determines sign of the order one parameter $r$. For the sake of concreteness, we consider the Standard Model gauge fields localised on a GUT $D_7$ brane in the context of F-theory \cite{beasley2009guts}.\footnote{Although there are no  realizations of the dark dimension scenario in string theory yet, this model could be close to a potential realization, where the standard model gauge fields are localised on a $D_7$ brane in the dark dimension.} Two length scales can be defined as
 \ba R_S &\equiv& {\cal V}(S)^{\sfrac{1}{4}},\\
     R_B &\equiv& {\cal V}(B)^{\sfrac{1}{6}},\ea
where $B$ is the base of the $CY_4$ and $S$ is the localization brane of the SM. The radius of dark dimension, the radius normal to $S$, is
\be R=R_B \Big(\frac{R_B}{R_S}\Big)^{\nu}, \ee
where for tubular geometries which describe dark dimension scenario $\nu\sim 2 $. After compactification to four dimensions we find the gauge kinetic function as
\be f\simeq \frac{R_S^4}{R_B^3}.\ee
Then, we find
\ba \frac{\partial f}{\partial R}=\frac{\partial f}{\partial R_B}\left(\frac{\partial R}{\partial R_B}\right)^{-1}+\frac{\partial f}{\partial R_S}\left(\frac{\partial R}{\partial R_S}\right)^{-1} 
=-\frac{(4+7\nu)}{\nu(1+\nu)}\left(\frac{R_S}{R_B}\right)^{4+\nu}<0,\ea
and consequently 
\be r<0. \ee
This is a generic property that the gauge kinetic function of a localised gauge theory increases when radius of dark dimension decreases. The gauge kinetic function of the gauge theory is proportional to its localization volume; the radius of the dark dimension increases as the ratio of the total radius to localization radius increases.  

\begin{figure}[t]
	\centering
	\includegraphics[width=0.4\linewidth]{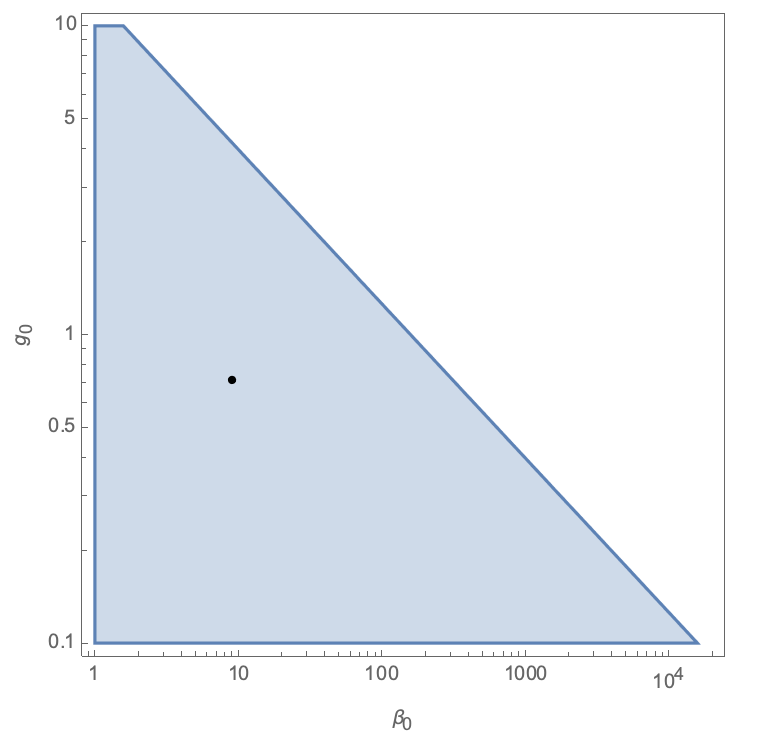}
	\caption{The gauge coupling versus the beta function coefficient. The blue region is the parameter space of Yang-Mills theories with satisfy \eqref{plotinequ}. The black dot is the Standard Model QCD.}
	\label{the bound}
\end{figure}

Finally, we find the condition that the FL bound is satisfied by a localized Yang-Mills theory in the dark dimension scenario.  Assume that along a path in the moduli space the dark dimension radius changes from $R_0$ to $R$ such that  $R_0>R$ or equivalently $\Lambda_0<\Lambda$. We recall that positive distance between the two points in the space of metrics implies $k>0$. Then given that $r$ and $k$ are negative and positive order one constants respectively, \eqref{cnd2i} implies that the following condition must be satisfied so that the FL bound holds in the vacuum with higher cosmological constant
\ba
\label{plotinequ}&&0<\beta_0\lesssim\frac{16 \pi^2}{g_0^2}, \ea
 where $g_0$ is the tree level gauge coupling at string scale. It runs with energy due to loop corrections and matches to the experimental values measured at low energy. In Figure \ref{the bound} we plot \eqref{plotinequ} for $g_0$ versus $\beta_0$ where the blue region is the parameter space of Yang-Mills theories that respect the FL bound. In this figure, We note that for a given gauge coupling at the string scale the beta function coefficient, which is fixed by the field content of the gauge theory, is bounded. For instance, if the beta function is dominated by the self interaction of gluons (which is the case for confining theories) then $\beta_0\simeq \sfrac{11}{3} N_c$ and  we find that the rank of the gauge group is bounded 
 \be N_c\lesssim 10^3. \ee  The black dot shows the Standard Model QCD with $\beta_0=9$ and $g_0^{\text{QCD}}=0.7$ at the unification scale which is confining today and is so in any other dS vacuum with higher cosmological constant.

\section{Conclusion}\label{sec7}
In this paper we studied dependence of gauge coupling on distance in the space of metric configurations in a theory of quantum gravity. We considered supersymmetric compactifications of heterotic, type IIA with $D6$ branes, type IIB with $D7/D3$ branes, type IIB with $D9/D5$ branes and argued that gauge coupling and confinement scale typically depend on distance in the metric space. In the (A)dS space, our results imply that confinement scale depends on the cosmological constant. We found the condition that the swampland FL bound for confinement scale of a gauge theory in dS space be realised in this framework; if the bound holds in a dS vacuum it also holds in a vacuum with higher cosmological constant, provided that stringy effects are stronger than radiative effects. Furthermore, we studied distance dependent confinement scale for gauge fields localised on a brane in dark dimension scenario. 
We argued that confinement scale increases as the radius of the dark dimension decreases. We also found that the FL bound is inherited to vacua with higher cosmological constants for Yang-Mills theories with $N_c\lesssim 10^3$.

In this work, we did not compute the exact distance dependence of the gauge kinetic function; we computed the leading order terms in the expansion in small changes of distance. The next step would be to go beyond this approximation, and find out how it might be related to other swampland conjectures such as the WGC. We postpone this to a future work. Finally, for the sake of concreteness and computational ability, we have considered supersymmetric compactifications. It would be interesting to look into realistic non-supersymmetric  examples. However, one expects that gauge kinetic functions depend on volume moduli and the dilaton in every compactification which consequently induces distance dependence. It will be studied in more detail in a future work. Furthermore, we have assumed the AdS distance conjecture as granted in this paper. However, it is interesting to explore the generalized distance conjecture for AdS within this specific context and investigate the possibility that open string moduli may alleviate the conformal problem. This is beyond the scope of this work and is left for further investigation in future work.

\paragraph*{Acknowledgements}
We would like to thank Alek Bedroya and José Calderón Infante for illuminating discussions. Special thanks go to Irene Valenzuela for fruitful discussions and comments on the draft. AM is thankful to the TH department of CERN for hospitality during the final stages of this work. MT thanks Perimeter Institute for warm hospitality.

\bibliographystyle{plain}

\end{document}